\documentclass{elsart}
\usepackage{graphicx}

\begin{document}

\begin{frontmatter}
\title{The operation threshold of a double barrier phonon laser}

\author[Cuba,UFF]{I. Camps \thanksref{CLAF}} 
\author[UERJ,UFF]{S. S. Makler} 

\address[Cuba]{Physics Faculty - IMRE, University of Havana, Cuba}
\address[UFF]{Instituto de F\'{\i}sica, Universidade Federal Fluminense,
Niter\'oi, RJ, Brazil} 
\address[UERJ]{Instituto de F\'{\i}sica, Universidade Estadual do Rio de
 Janeiro,RJ, Brazil}

\thanks[CLAF] {Partially supported by Latin American agency CLAF.}

\begin{abstract}
We make an adaptation of laser modelling equations to describe the behavior
of a phonon laser (saser). Our saser consists of an \textit{AlGaAs/GaAs} 
double barrier heterostructure designed to generate an intense 
beam of transversal acoustic ($TA$) phonons. To study our system, we 
begin with a Hamiltonian that describes the decay of primary 
longitudinal optical phonons ($LO_1$) into secondary ($LO_2$) and 
$TA$ ($LO_1\rightarrow LO_2 + TA$) and its inverse process 
(recombination). Using this Hamiltonian, a set of coupled equations of 
motion for the phonons is obtained. We also consider the interaction 
between the phonons and its reservoirs. These interactions are 
introduced in the equations of motion leading to a set of coupled 
Langevin equations. In order to obtain an expression to describe our 
saser we apply, in the Langevin equations, an adiabatic elimination of 
some variables of the subsystem. Following the method
above we obtain the value of the injection threshold for the operation 
of our phonon laser. At this threshold occurs a phase transition from 
a disordered to a coherent state. It is shown that it is not necessary a
big ``optical" pumping to get a sasing region.
\end{abstract}

\end{frontmatter}
\newpage

\section{Introduction}

Sasers are devices capable of producing beams of coherent phonons. 
Besides the system considered here, several other kinds of such devices 
were proposed and constructed
\cite{Fokker1,Fokker2,Prieur1,Prieur2,Zav1,Zav4,Watson}.

In recent works \cite{jpcm,surf,Tuyarot,Weber,bjp2,jpcm2} the kinetics 
and the dynamical properties of the double barrier phonon laser were 
discussed. In this paper we study the system from a point of view
close to that used in quantum optics to study the onset of the lasing
regime. In spite of the fact that the system is far from equilibrium and
thus concepts like temperature and chemical potential can not be
applied, the process is analogous to a phase transition
\cite{Haken-p1,Haken2}. This transition is not governed by
temperature but an energy flux  (in this case by the injection rate
of primary longitudinal optical $LO_1$ phonons). We were able to 
obtain the potential that describes the states of our system and 
consequently, the threshold value for the coherent transversal acoustic
($TA$) phonon emission. 
 
The paper is organized as follows. A description of the saser device is 
presented in section \ref{the system}. The multimode case is analyzed in
section \ref{multimode}. In section \ref{single mode} the potential 
that describes our system is obtained. Finally, in section 
\ref{results} we present our results and conclusions.
  
\section{The system}
\label{the system}
 
The phonon laser studied here consists of an \textit{AlGaAs/GaAs}
double-barrier heterostructure (DBH). This device was designed in such a
way that, for a small applied bias $V$, the difference 
$\Delta\varepsilon$ between the first excited level $\varepsilon _{1}$ 
and the ground state $\varepsilon _{0}$ in the well is less than the 
longitudinal optical ($LO_1$) phonon energy $\hbar \omega _{1}$. For a 
greater bias the resonant condition 
$\Delta\varepsilon\approx\hbar\omega _{1}$ is achieved and the 
electrons begin to decay to the ground state by emitting primary $LO_1$
phonons. For an $Al$ concentration greater than $0.25$ \cite{Juss} or 
$0.3$ \cite{Jacob}, these phonons are confined inside the well (they 
can be also absorbed by exciting electrons from $\varepsilon _{0}$ to 
$\varepsilon _{1}$).
 
The process described above, acts in parallel with the decay of primary 
$LO_1$ phonons due to anharmonicity. One of the products of this decay 
is a secondary longitudinal optical phonon ($LO_2$), the other is a 
$TA$ phonon \cite{Vall2}. The $LO_2$-$TA$ pair is produced by stimulated
emission. Therefore these $TA$ phonons could be coherent and form a 
beam that it is called saser by analogy with a laser.

The DBH saser beam could be applied to acoustic nanoscopy, 
phonoelectronics and phonolitography. This was discussed in more detail
in a previous article \cite{jpcm}. 

\begin{figure}[htbp]
\centering
\includegraphics[width=\textwidth]{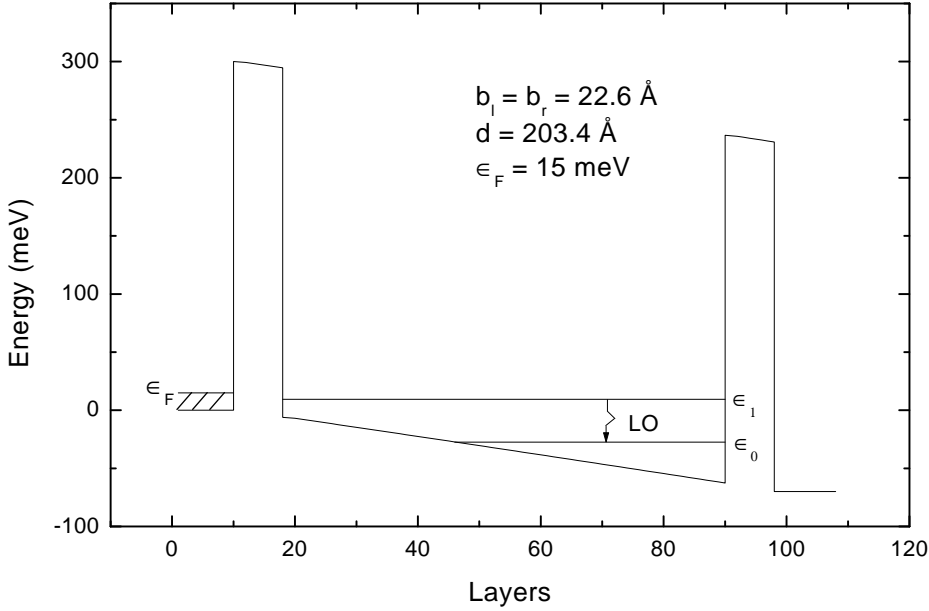}
\caption{Potential profile and energy levels at resonant condition. 
$b_l$, $b_r$, $d$ and $\epsilon_F$ are respectively the barriers and well widths 
and the Fermi level at the emitter.}
\label{profile-Fig1}
\end{figure}

The potential profile and the level positions at the resonant condition 
are shown in figure \ref{profile-Fig1}.

\section{The multimode case}
\label{multimode}

The fundamental mechanism in this saser is the generation of the $TA$ 
phonons due to the decay of primary $LO_1$ phonons. Therefore, the other
processes (i.e., emission-absorption of the $LO_1$ phonon through 
electron transitions and the $LO_2$ phonon decay) are secondary and 
are taken into account just as a result of the interaction of each 
phonon with its reservoir.

As the $TA$ phonon emission results from the decay of $LO_1$ phonons, 
they could be emitted in many modes, corresponding to the different 
values of its wave vector $\mathbf{q}_{3}$ parallel to the interface. 

The number $N_{\mathbf{q}_{3}}$ of wave vectors parallel to the 
interface can be estimated as $N_{\mathbf{q}_{3}}\sim S_0/a^2$, where 
$S_0$ is the device area and $a$ is the lattice parameter. For the 
values used here, $S_0=0.5\cdot 10^{-3}\ mm^{2}$ and $a=2.825$ \AA , 
$N_{\mathbf{q}_{3}}$ is of the order of $7\cdot 10^9$. The total number 
of emitted $TA$ phonons is $\sim 2\cdot 10^4$ \cite{jpcm2}, which means 
that, at the beginning of the phonon emission, the number of phonons per
wave vector $\mathbf{q}_{3}$ is $\sim 3\cdot 10^{-6}$. As the applied 
bias increases, the emission in a particular mode called 
$\mathbf{q}_{3}^0$ grow slowly until a certain threshold value 
$V = V_{th}$ is reached. For values of $V$ greater than $V_{th}$, 
the phonons in that mode are emitted in a coherent way 
implying that the population $n_{\mathbf{q}_{3}^0}$ is the only one 
macroscopically non zero. In this section, the possibility of 
macroscopically coexistence of more than one mode is studied.         
 
\subsection{The Hamiltonian}

The Hamiltonian that describes our system can be written as:

\begin{eqnarray}
\mathcal{H}=&&\hbar\omega _{1}\sum_{\mathbf{q}_{1}}b_{\mathbf{q}_{1}}^
{\dagger }b_{\mathbf{q}_{1}}+\hbar\omega _{2} \sum_{\mathbf{q}_{2}}
b_{\mathbf{q}_{2}}^{\dagger}b_{\mathbf{q}_{2}}+\hbar \omega _3
\sum_{\mathbf{q}_{3}}b_{\mathbf{q}_{3}}^{\dagger }b_{\mathbf{q}_{3}}+  
\nonumber \\
&&{\hbar \gamma }\sum_{\mathbf{q}_{1},\mathbf{q}_{2},\mathbf{q}_{3}}
\left( b_{\mathbf{q}_{1}}^{\dagger }b_{\mathbf{q}_{2}}b_{\mathbf{q}_{3}}
+b_{\mathbf{q}_{1}}b_{\mathbf{q}_{2}}^{\dagger }b_{\mathbf{q}_{3}}^
{\dagger }\right) \label{H_total},
\end{eqnarray}

where $\mathbf{q}_{1}$, $\mathbf{q}_{2}$ and $\mathbf{q}_{3}$ are 
respectively the $LO_{1}$, $LO_{2}$  and $TA$ wave vectors parallel to 
the interfaces, and must satisfy $\mathbf{q}_{1}=\mathbf{q}_{2}+
\mathbf{q}_{3}$, $\hbar \omega _i$ $(i=1,2,3)$ are the energies of each 
kind of phonons and $\gamma$ is the phonon-phonon interaction 
coefficient.

\subsubsection{The Langevin equations}
 
The equations of motion for the operators $n_{\mathbf{q}_{1}}$ 
($n_{\mathbf{q}_{i}}=b_{\mathbf{q}_{i}}^{\dagger }b_{\mathbf{q}_{i}}$), 
$b_{\mathbf{q}_{2}}$, $b_{\mathbf{q}_{3}}$ are obtained from the 
Hamiltonian (\ref{H_total}). We remark that the saser system is coupled 
to reservoirs. The $LO_{1}$ phonons are coupled through the emission and 
absorption processes to the electron system. In the case of the $LO_2$ 
phonons, they are coupled to the other phonon modes to which they decay,
whereas the $TA$ phonons are connected to the other $TA$ phonon modes 
outside the well. Thus the total Hamiltonian includes, besides the 
terms shown in equation (\ref{H_total}), the interaction with the 
reservoirs. By using the methods described in references 
\cite{Cohen1,Weiss}, the reservoir coordinates can be eliminated leading
to the following Langevin equations

\begin{eqnarray}
\label{langevin1}
\frac{dn_{\mathbf{q}_{1}}}{dt} &=&-i\gamma \sum\limits_{\mathbf{q}_{2},
\mathbf{q}_{3}}{\left\{ {b_{\mathbf{q}_{1}}^{\dagger }b_{\mathbf{q}_{2}}
b_{\mathbf{q}_{3}}-b_{\mathbf{q}_{1}}b_{\mathbf{q}_{2}}^{\dagger}
b_{\mathbf{q}_{3}}^{\dagger }}\right\} +G_{\mathbf{q}_{1}}-\Gamma _{1}
n_{\mathbf{q}_{1}}}+ \hat F_1,  \label{langevin11} \\
\frac{db{_{\mathbf{q}_{2}}}}{dt} &=&-(i\omega _2 +\kappa _{2})
b_{\mathbf{q}_{2}}-i\gamma \sum\limits_{\mathbf{q}_{1},\mathbf{q}_{3} }
{b_{\mathbf{q}_{1}}b_{\mathbf{q}_{3}}^{\dagger }}+\hat F_2,  
\label{langevin12} \\
\frac{db_{{\mathbf{q}_{3}}}}{dt}&=&-(i\omega _3 +
\kappa _{\mathbf{q}_{3}})b_{\mathbf{q}_{3}}-i\gamma
\sum\limits_{\mathbf{q}_{1},
\mathbf{q}_{2}}{b_{\mathbf{q}_{1}}b_{\mathbf{q}_{2}}^{\dagger }}+
\hat F_{3},\label{langevin13} 
\end{eqnarray}
where $G_{\mathbf{q} _{1}}$ and $\Gamma _{1}$ are the emission and 
absorption rates for the $LO_1$ phonons due the interaction with its 
reservoir (i.e., the electrons), $\kappa _{2}$ is the decay rate of 
$LO_2$ phonons, $\kappa _{\mathbf{q}_{3}}$ is the escape rate of the 
$TA$ phonons and $\hat F_i\, (i=1,2,3)$ are the fluctuation forces. The 
$TA$ phonons escape rate is different for each mode, thus depending on 
${\mathbf{q}_{3}}$. As we are not interested in studying the statistical 
properties of the saser, in the following the fluctuation forces will 
not be considered.

In order to obtain only one equation for the $TA$ phonons, we make the 
adiabatic approximation in equations (\ref{langevin11}) and 
(\ref{langevin12}). First we replace $b_{{\bf q}_{1}}\left( 
t \right)=\tilde b_{{\bf q}_{1}}\left( t \right)e^{ - i\omega _1 t}$ and
$b_{{\bf q}_{3}}\left( t \right)=\tilde b_{{\bf q}_{3}}\left( t \right)
e^{ - i\omega _3 t}$ in (\ref{langevin12}). Secondly, we integrate 
(\ref{langevin12}) explicitly to get

\begin{equation}
b_{{\bf q}_{2}}=-i\gamma \sum_{{\bf q}_{1},{\bf q}_{3}}
\int \limits _{-\infty}^{t}e^{-\left[i\left( \omega _1  - \omega _2  - 
\omega _3 \right)+\kappa _{2} \right] \left( {t-\tau }\right) }\left( 
{\tilde b_{{\bf q}_{1}}\tilde b_{{\bf q}_{3}}^{\dagger }}\right) _{\tau}
d\tau.
\label{integra-b2} 
\end{equation} 
Assuming that the relaxation time of the $b_{{\bf q}_{2}}$ phonons is 
much smaller than the relaxation of $\tilde b_{{\bf q}_{1}}$ and 
$\tilde b_{{\bf q}_{3}}^\dag$, we can take $\tilde b_{{\bf q}_{1}}
\tilde b_{{\bf q}_{3}}^{\dagger}$ out of the integral in equation 
(\ref{integra-b2}) obtaining

\begin{equation}
b_{\mathbf{q}_{2}}  =  - i\gamma \sum\limits _{\mathbf{q}_{1} ,
\mathbf{q}_{3}} \frac{{e^{ - i\left( {\omega _1  - \omega _3 } \right)t}
}}{{ - i\nu + \kappa _{2}}} \tilde b_{\mathbf{q}_{1}} 
\tilde b_{\mathbf{q}_{3}}^\dag, \label{b2}  
\end{equation}
where $\nu = {\omega _1  - \omega _2  - \omega _3}$.

In a similar way, we obtain for $n_{\mathbf{q}_{1}}$ 

\begin{equation}
n_{\mathbf{q}_{1}} = \frac{G_{\mathbf{q}_{1}}}{\Gamma _1}\left( %
1 - \frac{\Omega}{\Gamma _1 }\sum\limits _{\mathbf{q}_{3}} 
{n_{\mathbf{q}_{3}}} \right), \label{n1} 
\end{equation}
where 
\[\Omega = \frac{2\gamma ^2 \kappa _2 }{\nu^2 + \kappa _2^2 }.\]
Finally, we replace (\ref{b2}) and (\ref{n1}) in (\ref{langevin13}) and 
we get, after some manipulations, an equation for the $TA$ phonon 
populations

\begin{equation}
\frac{dn_{\mathbf{q}_{3}}}{dt} = (\Omega n_{1}-2\kappa _{\mathbf{q}_{3}}
)n_{\mathbf{q}_{3}},
\label{dn3} 
\end{equation}
where $n_{1} = \sum\limits_{\mathbf{q}_{1}}{n_{\mathbf{q}_{1}}}$.

Considering as an example only two modes, $\mathbf{q}_{3}^0$ and 
$\mathbf{q}_{3}^1$, we get for the stationary case of equation 
(\ref{dn3}):

\begin{eqnarray}
\label{two-modes}
n_{3}^{0}\left( \Omega n_{1} - 2\kappa _{3}^{0} \right) = 0, 
\label{mode1}\\ 
n_{3}^{1}\left( \Omega n_{1} - 2\kappa _{3}^{1} \right) = 0. 
\label{mode2}
\end{eqnarray}

If both modes would be present, ($n_{3}^{0}\neq 0$ and 
$n_{3}^{1}\neq 0$) both parenthesis in (\ref{mode1}) and (\ref{mode2}) 
would be zero. This is a contradiction (because 
$\kappa _{3}^0\neq\kappa _{3}^1$). This contradiction only can be solved 
if just one mode is present and the other one has died out. Therefore, 
only a single mode, the one with largest lifetime inside the well and 
closest to resonance, survives. The former analysis also can be done 
quite rigorously for many modes.

When the $LO_1$ decay, the $TA$ phonons are emitted in the [$111$] 
direction \cite{Vall}. The device is grown in this 
direction, such that the $TA$ phonon beam will make multiple reflections
between the walls of the well, implying in large lifetime for these 
phonons. Indeed, for a small applied bias, the $TA$ phonons are emitted 
with any wave vector $\mathbf{q}_{3}$. In this regime, the device works
as a sound emitting diode (SED). With a further increase in the 
applied potential, when a certain value of $V$ is reached, the $TA$ 
phonon distribution on $\mathbf{q}_{3}$ become more and more sharp 
giving place to a phase transition. In this case $n_{3}^0$ grows 
suddenly from about $10^{-6}$ to about $10^{4}$. Thus, a great number of
$TA$ phonons would have only one mode: $\mathbf{q}_{3}^0=0$ (which 
imply $\mathbf{q}_1 = \mathbf{q}_2$). This mode will survive and slave 
the others, making the system to self-organize and emit just in this 
mode.

\section{The single mode case}
\label{single mode}

In order to get the threshold for which $n_{3}^0$ begins to be 
macroscopically non zero, we must first obtain an equation for the 
amplitude of the single mode $b_{3}^0$ that slaves the system. To do 
that, we make the same approximations as described in the previous 
section. The equation obtained for $b_{3}^0$ is

\begin{equation}
\frac{{db_3 }}{{dt}} = \left( {\frac{{\gamma ^2 }}{{\Gamma _1 
\kappa _2 }}G_1  - \kappa _3 } \right)b_3^0  - 2G_1 \left( {\frac{{
\gamma ^2 }}{{\Gamma _1 \kappa _2 }}} \right)^2 {b_3^0}^\dag  b_3^0 
b_3^0 \label{db3},
\end{equation}
where $G_{1} = \sum\limits_{\mathbf{q}_{1}}{G_{\mathbf{q}_{1}}}$.

The right-hand side of (\ref{db3}) can be obtained from the potential:

\begin{equation}
U_c \left({b_3^0}\right) = \alpha\left|{b_3^0}\right|^2 
+ \beta\left| {b_3^0 } \right|^4,
\label{potential-b3} 
\end{equation}
with
\[
\alpha = {\kappa _3-\frac{{\gamma ^2}}{{\Gamma _1\kappa _2}}G_1},\quad 
\beta = G_1\left({\frac{{\gamma ^2}}{{\Gamma _1\kappa _2}}}\right)^2,
\]
such that

\begin{equation}
\frac{{db_3^0 }}{{dt}} =  - \frac{{dU_c \left({b_3^0}
\right)}}{{db_3^\dag}}.
\end{equation}
The expressions for the parameters $G_1$ and $\Gamma _1$ are 
\cite{jpcm2}
 
\[ \Gamma _1 = w \left(N_0 - N_1 \right),~~~~G_1 = w N_1~. \]
where $w$ is the electron transition rate and $N_0$ and $N_1$ are 
respectively the electron populations of the fundamental and excited 
levels. 

The shape of the potential (\ref{potential-b3}) is similar to that 
obtained in the Ginzburg-Landau phase transition theory. 
For $\alpha>0$ the potential has a minimum at $n_{3}^0=0$ 
($n_{3}^0=\left|{b_3^0 }\right|^2$). In the case of $\alpha<0$, the 
minimum is obtained for $n_{3}^0\neq 0$. The transition from $\alpha>0$ 
to $\alpha<0$ is due to the competition between processes of gain,
with a rate $G_1 {\frac{{\gamma ^2 }}{{\Gamma _1 \kappa _2 }}}$ 
and losses, with a rate $\kappa _3$. The threshold is attained when 
$\alpha=0$, i.e.,
$G_1 = {\frac{{\Gamma _1 \kappa _2 \kappa _3}}{{\gamma ^2 }}}$.  

\section{Results and Conclusions}
\label{results}

The parameters used in our calculation are: the escape rate of $TA$ 
phonons $\kappa _3^0=0.05\; ps^{-1}$, estimated from its group velocity 
near the $L$ point of the Brillouin zone, the decay rate
of the $LO_2$ phonons $\kappa _2=20\; ps^{-1}$ (by assuming a small 
stimulated decay) and the phonon-phonon interaction coefficient 
$\gamma=0.5\; ps^{-1}$ calculated in reference \cite{tobe} following
the work of Klemens \cite{Klemens}. The decay rate $\gamma _0$ obtained
from this value is in agreement with the experimental result of 
Vall\'ee \cite{Vall2}. The values of $G_1$ and $\Gamma _1$ depend on 
the $LO_1$ phonon emission rate $w$ and on the electron 
populations $N_0$ and $N_1$ that were calculated in reference 
\cite{jpcm2} by solving the kinetic equations. 
   
The total generation rate of the $LO_1$ phonons $G_1$, depends on the 
applied bias $V$ (i.e., it turns out from electron transitions that 
are $V$ dependent \cite{jpcm2}). Thus, the applied bias is the 
external parameter that control the intensity of the saser emission.

\begin{figure}[htbp]
\centering
\includegraphics[width=\textwidth]{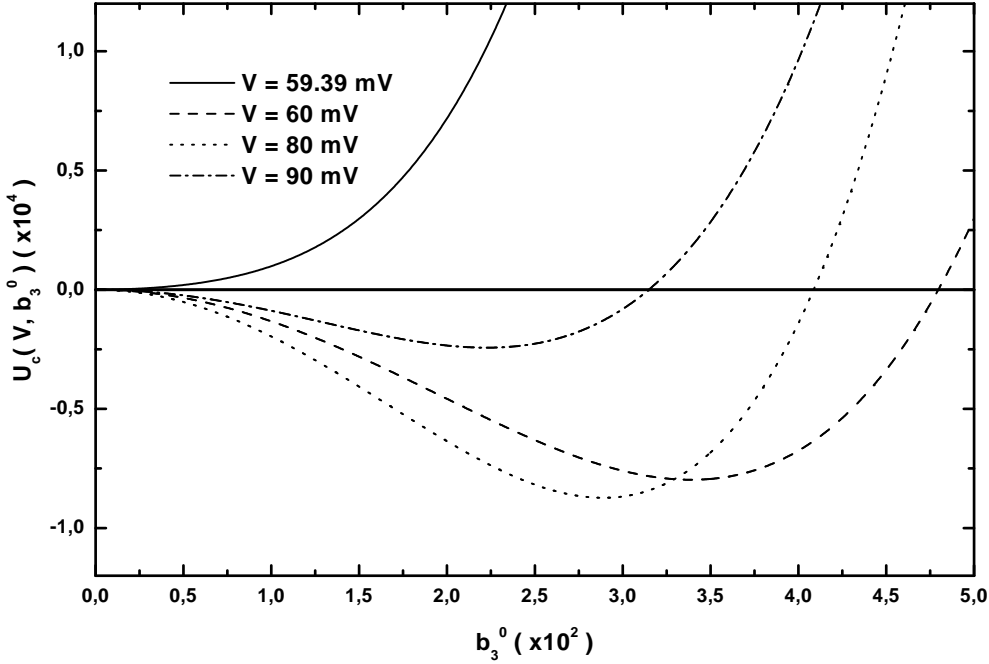}
\caption{The potential $U_c$ as a function of the $TA$ phonon amplitude
for different values of the applied bias.}
\label{potential-Fig2}
\end{figure}

In figure \ref{potential-Fig2}, the potential (\ref{potential-b3}) is 
plotted as a function of the $TA$ phonon amplitude for different values 
of the applied bias $V$. As it can be seen, for $V > V_{th}=59.4\; mV$ 
this potential has different minima for different values of $V$. These  
minima represent the stable points $b_3^s$ of the system.
\begin{figure}[htbp]
\centering
\includegraphics[width=\textwidth]{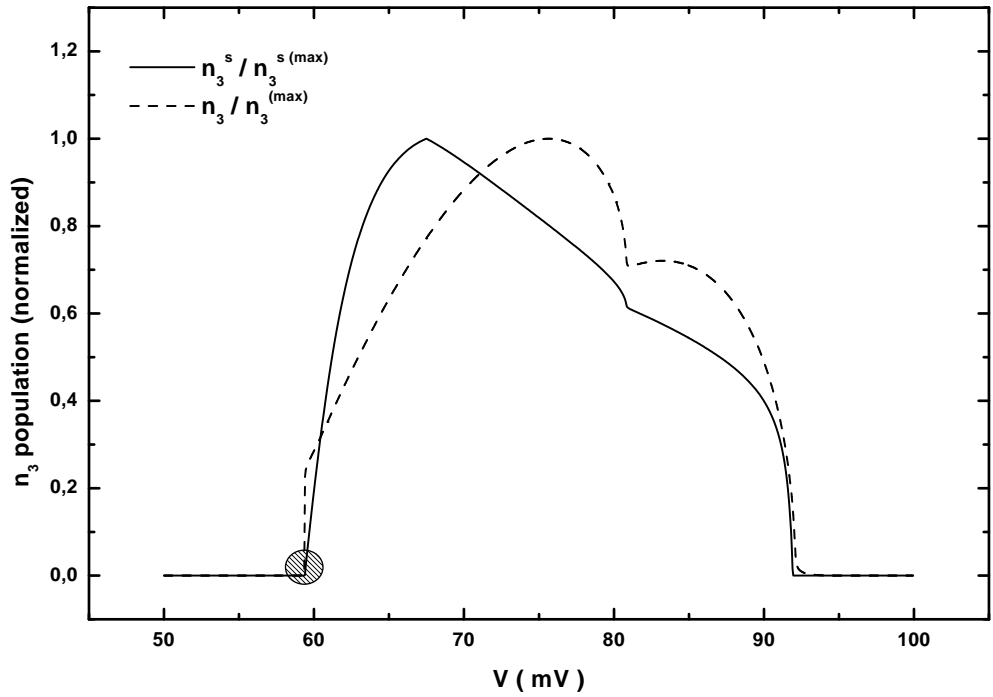}
\caption{The solid line shows the result for the
\textit{coherent} mode population obtained here (the minima of the 
previous figure), whereas the dashed line shows the \textit{total} $TA$ 
phonon population obtained from the kinetic equations \cite{jpcm2}.}
\label{n3-Fig3}
\end{figure}

In figure \ref{n3-Fig3} the stable points are plotted as function of $V$. 
The solid line represents the $n_3^s$ calculated in this work. The 
dashed line is the $TA$ phonon population calculated in a previous work
from the kinetic equations \cite{jpcm2}. 
\begin{figure}[htbp]
\centering
\includegraphics[width=\textwidth]{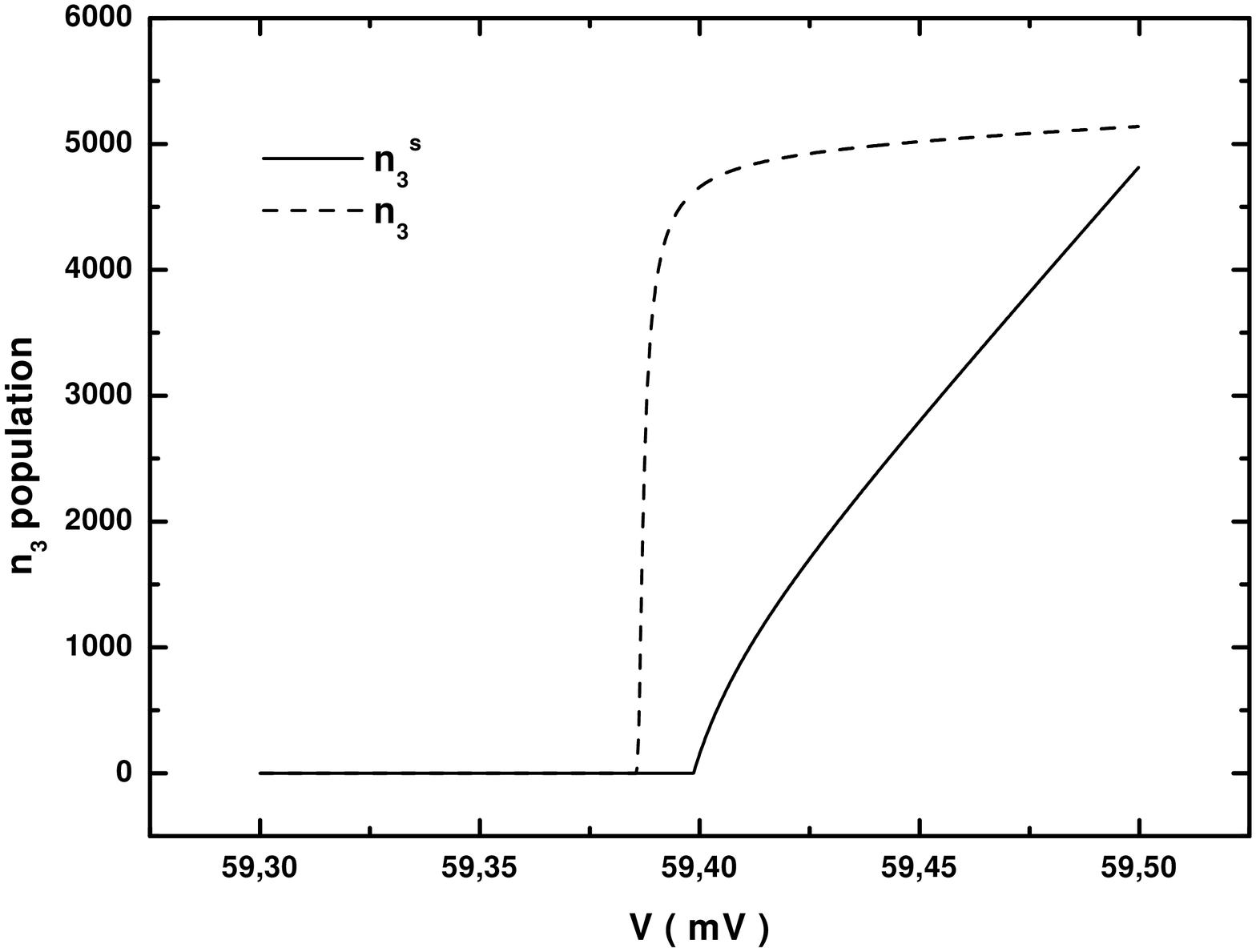}
\caption{The dashed region in figure 3 is shown in detail. We can see that
the $TA$ emission process begins incoherent. It becomes coherent when 
the injection rate of phonons $LO_1$ is $G_1 \geq \frac{ \Gamma _1 
\kappa _2 \kappa_3}{\gamma^2}$.}
\label{comparison-Fig4}
\end{figure}

Figure \ref{comparison-Fig4} shows a comparison in the region where the 
$TA$ generation begins (dashed region in figure \ref{n3-Fig3}). From 
this comparison two conclusions arise. First, as it can be seen (solid
line), the $TA$ coherent emission does not start until a certain 
threshold of the applied bias is reached. At $V_{th}$ the $LO_1$ 
emission rate $G_1$ is of the order of $279\; ps^{-1}$ and the total $TA$
phonon population $n_3$ is $\sim 4658$, whereas $n_3^s=0$.
The $TA$ phonons generated for a bias greater than $V_{th}$ are 
coherent. Secondly, our previous calculations \cite{jpcm2},
even without considering the self-organization process, gave results 
that agree quite well with those presented here. This results show 
that the coherent emission occurs immediately after the condition of 
$LO_1$ phonon emission is reached. That means that it is not necessary a
big ``optical" pumping to get a sasing region.
In summary, we show in this work that the $TA$ phonons, produced by stimulated 
emission, constitute a coherent phonon beam in the parameter region discussed above.


\newpage
\begin{thebibliography}{10}

\bibitem{Fokker1}
P.~A. Fokker, J.~I. Dijkhuis, and H.~W. de~Wijn,
\newblock ``Stimulated emission of phonons in an acoustical cavity,''
\newblock {\em Phys. Rev. B}, vol. 55, no. 5, pp. 2925 -- 2933, Feb. 1997.

\bibitem{Fokker2}
P.~A. Fokker, R.~S. Meltzer, Y.~P. Wang, J.~I. Dijkhuis, and H.~W. de~Wijn,
\newblock ``Suppression of stimulated phonon emission in ruby by a
  magnetic-field gradient,''
\newblock {\em Phys. Rev. B}, vol. 55, no. 5, pp. 2934 -- 2937, Feb. 1997.

\bibitem{Prieur1}
J.~Y. Prieur, R.~H\"ohler, J.~Joffrin, and M.~Devaud,
\newblock ``Sound amplification by stimulated emission of radiation in an
  amorphous compound,''
\newblock {\em Europhys Lett.}, vol. 24, no. 5, pp. 409 -- 414, Nov. 1993.

\bibitem{Prieur2}
J.~Y. Prieur, M.~Devaud, J.~Joffrin, C.~Barre, and M.~Stenger,
\newblock ``Sound amplification by stimulated emission of phonons using
  two-level systems in glasses,''
\newblock {\em Physica B}, vol. 220, pp. 235 -- 238, Apr. 1996.

\bibitem{Zav1}
S.~T. Zavtrak,
\newblock ``Acoustic laser with dispersed particles as an analog of a
  free-electron laser,''
\newblock {\em Phys. Rev. E}, vol. 51, no. 3, pp. 2480 -- 2484, Mar. 1995.

\bibitem{Zav4}
S.~T. Zavtrak,
\newblock ``Acoustical laser with mechanical pumping,''
\newblock {\em J. Acoust. Soc. Amer.}, vol. 99, no. 2, pp. 730 -- 733, Feb.
  1996.

\bibitem{Watson}
Andrew Watson,
\newblock ``Pump up the volume,''
\newblock {\em New Scientist}, pp. 37 -- 40, Mar. 1999.

\bibitem{jpcm}
S.~S. Makler, M.~I. Vasilevskiy, E.~V. Anda, D.~E. Tuyarot, J.~Weberszpil, and
  H.~M. Pastawski,
\newblock ``A source of terahertz coherent phonons,''
\newblock {\em J. of Phys. Cond. Matt.}, vol. 10, no. 26, pp. 5905 -- 5921,
  Jul. 1998.

\bibitem{surf}
S.~S. Makler, D.~E. Tuyarot, E.~V. Anda, and M.~I. Vasilevskiy,
\newblock ``Ultra-high-frequency coherent sound generation in resonant
  tunneling,''
\newblock {\em Surf. Sci.}, vol. 362, no. 1--3, pp. 239 -- 242, 1996.

\bibitem{Tuyarot}
D.~E. Tuyarot, S.~S. Makler, E.~V. Anda, and M.~I. Vasilevskiy,
\newblock ``Double-barrier coherent sound generator: a new device,''
\newblock {\em Superlatt. Microstruct.}, vol. 22, no. 3, pp. 427 -- 430, 1997.

\bibitem{Weber}
J.~Weberszpil, S.~S. Makler, E.~V. Anda, and M.~I. Vasilevskiy,
\newblock ``A many body study of the {SASER} dynamics,''
\newblock {\em Microelectron. Eng.}, vol. 43 -- 44, pp. 471 -- 479, Aug. 1998.

\bibitem{bjp2}
I.~Camps, S.~S. Makler, and E.~V. Anda,
\newblock ``A quantum formalism for a terahertz acoustic laser,''
\newblock {\em Braz. J. of Phys.}, vol. 29, no. 4, pp. 694 -- 701, 1999.

\bibitem{jpcm2}
S.~S. Makler, I.~Camps, J.~Weberszpil, and D.~E. Tuyarot,
\newblock ``A double-barrier heterostructure generator of terahertz phonons:
  many-body effects,''
\newblock {\em J. of Phys. Cond. Matt.}, vol. 12, no. 13, pp. 3149 -- 3172,
  Apr. 2000.

\bibitem{Haken-p1}
H.~Haken,
\newblock ``Cooperative phenomena in systems far from thermal equilibrium and
  in nonphysical systems,''
\newblock {\em Rev. Mod. Phys.}, vol. 47, no. 1, pp. 67 -- 121, Jan. 1975.

\bibitem{Haken2}
H.~Haken,
\newblock {\em Synergetics. An Introduction. Nonequilibrium Phase
  Transitions and Self-Organization in Physics, Chemistry and
  Biology},
\newblock Springer Series in Synergetics. Springer-Verlag, Berlin, third
  edition, 1983.

\bibitem{Juss}
B.~Jusserand, F.~Mallot, J.~M. Moison, and G.~Leroux,
\newblock ``Atomic-scale roughness of {GaAs/AlAs} interfaces. {A
  Raman}-scattering study of asymmetrical short-period superlattices,''
\newblock {\em Appl. Phys. Lett.}, vol. 57, no. 6, pp. 560 -- 562, Aug. 1990.

\bibitem{Jacob}
J.~M. Jacob, D.~M. Kim, A.~Bouchalkha, J.~J. Sony, J.~F. Klem, H.~Hou, C.~W.
  Tu, and H.~Morko\c{c},
\newblock ``Spatial characteristic of {GaAs}, {GaAs}-like, and {AlAs}-like {LO}
  phonons in {GaAs}-{$Al_{x}Ga_{1-x}As$} superlattices. {T}he strong
  x-dependence,''
\newblock {\em Sol. State Comm.}, vol. 91, no. 9, pp. 721 -- 724, Sep. 1994.

\bibitem{Vall2}
F.~Vall\'{e}e and F.~Bogani,
\newblock ``Coherent time-resolve investigation of {LO}-phonon dynamics in
  {GaAs},''
\newblock {\em Phys. Rev. B}, vol. 43, no. 14, pp. 12 049 --12 052, May 1991.

\bibitem{Cohen1}
Claude Cohen-Tannoudji, Jacques Dupont-Roc, and Gilbert Grynberg,
\newblock {\em Atom-Photon Interactions},
\newblock John Wiley \& Sons, Inc., New York, 1992.

\bibitem{Weiss}
Ulrich Weiss,
\newblock {\em Quantum dissipative systems}, vol.~10 of {\em Series in Modern
  Condensed Matter Physics},
\newblock World Scientific Publishing, P O Box 128, Farrer Road, Singapore
  912805, second edition, 1999.

\bibitem{Vall}
F.~Vall\'{e}e,
\newblock ``Time resolved investigation of coherent {LO}-phonon relaxation in
  {III-V} semiconductors,''
\newblock {\em Phys. Rev. B}, vol. 49, no. 4, pp. 2460 -- 2468, Jan. 1994.

\bibitem{tobe}
I.~Camps and S.~Makler,
\newblock ``The terahertz phonon laser: a full quantum treatment,''
\newblock submitted for publication.

\bibitem{Klemens}
P.~G. Klemens,
\newblock ``Anharmonic decay of optical phonons,''
\newblock {\em Phys. Rev.}, vol. 148, no. 2, pp. 845 -- 848, Aug. 1966.

\end{thebibliography}
\end{document}